\documentclass[prl,superscriptaddress,citeautoscript,twocolumn]{revtex4-1}

\usepackage{graphicx}
\usepackage{amsmath,amssymb}
\usepackage{color}

\begin{document}

\title{Dissipative polarization domain walls in a passive driven Kerr resonator}

\author{Bruno Garbin}
\affiliation{Physics Department, The University of Auckland, Private Bag 92019, Auckland 1142, New Zealand}
\affiliation{The Dodd-Walls Centre for Photonic and Quantum Technologies, Dunedin, New Zealand}
\affiliation{Universit\'e Paris-Saclay, CNRS, Centre de Nanosciences et de Nanotechnologies, 91120, Palaiseau, France}

\author{Julien Fatome}
\affiliation{Physics Department, The University of Auckland, Private Bag 92019, Auckland 1142, New Zealand}
\affiliation{The Dodd-Walls Centre for Photonic and Quantum Technologies, Dunedin, New Zealand}
\affiliation{Laboratoire Interdisciplinaire Carnot de Bourgogne (ICB), UMR 6303 CNRS, Universit\'e de Bourgogne
             Franche-Comt\'e, 9 Avenue Alain Savary, BP 47870, F-21078 Dijon, France}

\author{Gian-Luca~Oppo}
\affiliation{SUPA and Department of Physics, University of Strathclyde, Glasgow G4 0NG, Scotland, European Union}

\author{Miro~Erkintalo}
\author{Stuart G. Murdoch}
\author{St\'ephane Coen}
\email{s.coen@auckland.ac.nz}
\affiliation{Physics Department, The University of Auckland, Private Bag 92019, Auckland 1142, New Zealand}
\affiliation{The Dodd-Walls Centre for Photonic and Quantum Technologies, Dunedin, New Zealand}

\begin{abstract}
   \noindent Using a passive driven nonlinear optical fiber ring resonator, we report the experimental realization
   of dissipative polarization domain walls. The domain walls arise through a symmetry breaking bifurcation and
   consist of temporally localized structures where the amplitudes of the two polarization modes of the resonator
   interchange, segregating domains of orthogonal polarization states. We show that dissipative polarization domain
   walls can persist in the resonator without changing shape. We also demonstrate on-demand excitation, as well as
   pinning of domain walls at specific positions for arbitrary long times. Our results could prove useful for the
   analog simulation of ubiquitous domain-wall related phenomena, and pave the way to an all-optical buffer adapted
   to the transmission of topological bits.
\end{abstract}

\maketitle

\noindent Domain walls (DWs) are self-localized kink-type topological defects that connect two stable states of a
physical system. They usually form in the presence of a spontaneous symmetry breaking bifurcation
\cite{golubitsky_singularities_1988}, and are found in a variety of contexts, including magnetism
\cite{weiss_lhypothese_1907}, hydrodynamics \cite{tsitoura_phase_2018}, biology \cite{reichl_modern_2016},
Bose-Einstein condensates \cite{coen_domain_2001, *stamper-kurn_spinor_2013}, and string theory
\cite{weinberg_quantum_1995}. The paradigmatic examples are the interfaces that separate domains with distinct
magnetization in ferromagnetic materials \cite{weiss_lhypothese_1907, parpia_direct_1983, *unguris_observation_1991},
whose unique properties are exploited in modern spintronics devices to store or even transfer information
\cite{allwood_magnetic_2005, *parkin_magnetic_2008, *currivan-incorvia_logic_2016}. Additionally, DWs are central to
numerous phase transitions in condensed matter and quantum physics \cite{reichl_modern_2016, dutta_quantum_2015}.

DWs are also known to manifest themselves in optical systems. In this context, the terminology was first used to
describe stationary spatial distributions of light arising from the pure nonlinear (Kerr) interactions of
counter-propagating beams \cite{zakharov_polarization_1987} (and reported experimentally
in~\cite{pitois_polarization_1998}). Subsequently, Haelterman and Sheppard introduced the concept of DW
\emph{solitons} by describing vector, kink-type propagating structures, segregating homogeneous domains of orthogonal
polarization states, and that resist diffractive (transverse) or dispersive (temporal) spreading in Kerr media
\cite{haelterman_pdw_1994, *sheppard_polarization-domain_1994}. Referred to as polarization DWs (or PDWs), these
structures have only recently been convincingly observed experimentally by Gilles et al --- more than two decades
after their theoretical description --- in the single-pass, \emph{conservative}, propagation configuration of a
normally dispersive single-mode optical fiber \cite{gilles_polarization_2017}. Remarkably, this experiment has
demonstrated the potential of temporal PDWs for transmission of topological bits, robust to noise and nonlinear
impairments, as originally foreseen~\cite{haelterman_polarisation_1994, *haelterman_dual-frequency_1995}.

Here we extend the work of Gilles et al~\cite{gilles_polarization_2017} by implementing all-optical storage of
temporal PDWs. This is obtained by taking advantage of spontaneous symmetry breaking in an externally-driven passive
Kerr ring resonator, enabling recirculation of PDWs \cite{fatome_polarization_2016}. To date, such \emph{dissipative}
PDWs have not been observed experimentally. Our demonstration constitutes a key technology in supporting potential
topologically-robust transmissions. It could also pave the way to real-time, stochastic, room temperature analog
simulations of DW-related solid-state physics phenomena not easily observable in other settings
\cite{oppo_characterization_2001, skryabin_perturbation_2001, rabbiosi_suppression_2002}. Finally, dissipative PDWs
are also related to phase DWs predicted in the transverse structure of optical parametric~\cite{trillo_stable_1997,
*oppo_domain_1999} and four-wave mixing~\cite{taranenko_pattern_1998, *esteban-martin_controlled_2005} oscillators.
As the dynamics of these systems can be used to estimate the ground state of the Ising Hamiltonian
\cite{wang_coherent_2013, *marandi_network_2014, inagaki_coherent_2016}, domains of orthogonal polarizations
segregated by dissipative PDWs could be associated with different spin states and provide a new route to solve
complex optimization problems. We note that hints of dissipative PDW ``complexes'' have been reported in fiber
lasers, but these observations have proved hard to interpret \cite{williams_fast_1997, zhang_observation_2009,
lecaplain_polarization-domain-wall_2013}. In contrast, the results we present in this Letter provide the first clear
signature of isolated dissipative temporal PDWs.

We start by describing theoretically how dissipative PDWs emerge in our system. We consider a dispersive passive ring
resonator that is externally and coherently driven by a continuous-wave (cw) light beam and that exhibits a Kerr
nonlinearity. The intracavity field is described in terms of the complex amplitudes, $E_{1,2}$, of the two orthogonal
polarization modes of the resonator. In the mean-field limit, the temporal evolution of these two modal amplitudes
can be described by normalized coupled Lugiato-Lefever equations as \cite{lugiato_spatial_1987,
haelterman_polarization_1994, averlant_coexistence_2017, nielsen_coexistence_2019},
\begin{multline}
 \frac{\partial E_{1,2}}{\partial t} = \biggl[-1 + i\left(|E_{1,2}|^2+B|E_{2,1}|^2-\Delta\right) \biggr.
     \\ \biggl. - i \frac{\partial^2 }{\partial \tau^2}\biggr] E_{1,2} + \sqrt{X/2}\,.
  \label{eq:coupledLLEs}
\end{multline}
The terms on the right hand side of these equations describe respectively cavity losses, self- and cross-phase
modulation, the cavity phase detuning, chromatic dispersion (taken as normal, to avoid modulational instabilities of
cw stationary states of the resonator \cite{haelterman_polarization_1994}), and external driving. We consider here
identical detuning and driving strength for the two modes. $t$ represents a slow time over which the evolution of the
intracavity field takes place, at the scale of the cavity photon lifetime, while $\tau$ is a fast-time that describes
the temporal structure of the field along the round trip of the resonator. $B$ is the cross-phase modulation
coefficient. In optical fibers (as used in our experiments discussed below), $B$ can be as large as~2 for circularly
polarized modes~\cite{agrawal_nonlinear_2013}, but PDWs continue to exist so long as $B>1$~\cite{haelterman_pdw_1994,
*sheppard_polarization-domain_1994}. Finally, $\Delta$ is the detuning parameter, which measures the separation
between the driving laser frequency and the nearest cavity resonance in terms of the cavity half-linewidth, and $X$
is the normalized total driving power.

\begin{figure}
  \centerline{\includegraphics[width=\columnwidth]{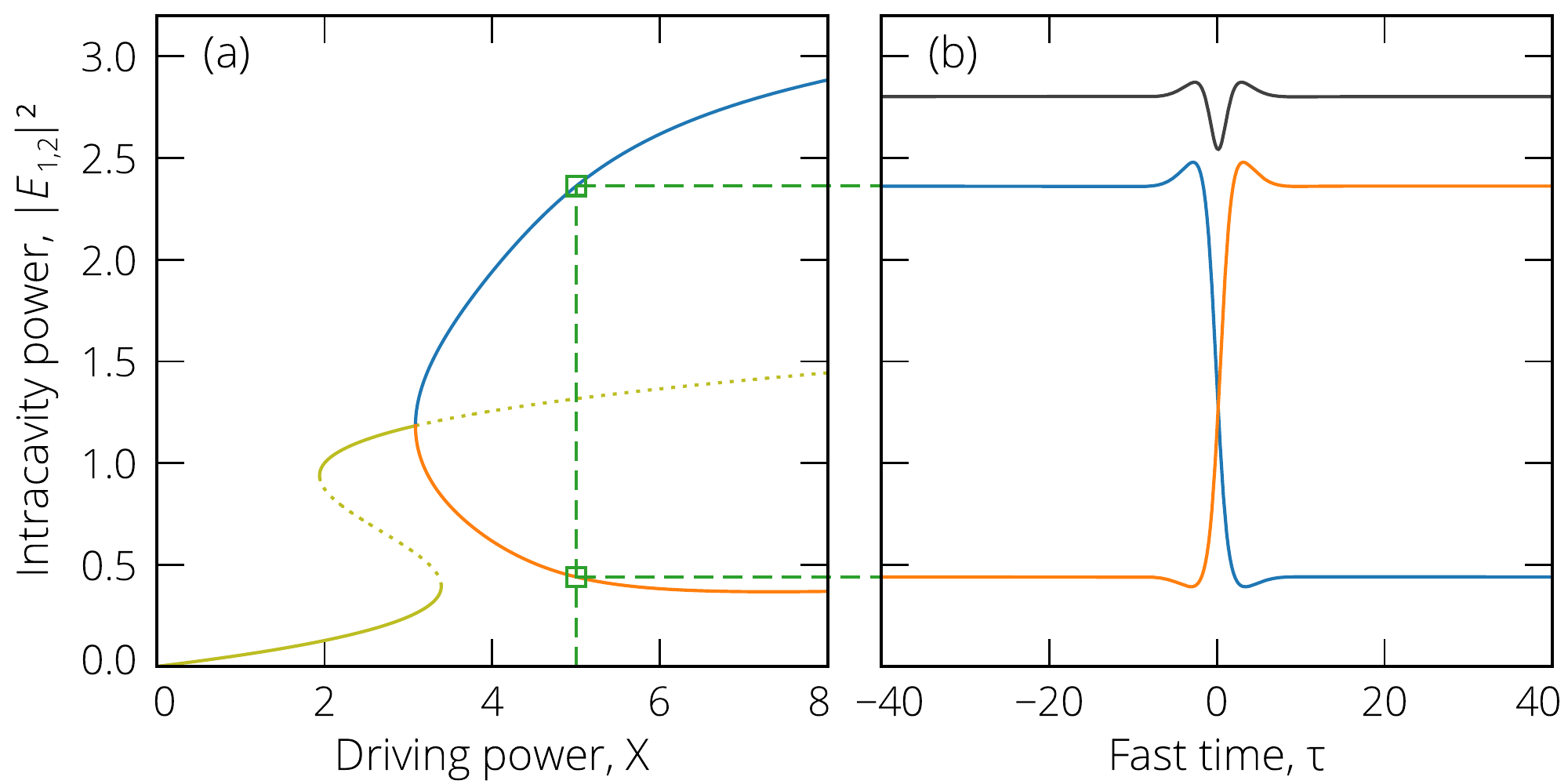}}
  \caption{Numerical illustration of the polarization SSB and associated PDWs for $\Delta=3$ and $B=2$ as described
    by Eqs.~(\ref{eq:coupledLLEs}). (a) Stationary cw solutions, $|E_1|^2$ and $|E_2|^2$, versus driving power~$X$.
    The yellow curve represents symmetric solutions ($E_1 = E_2$; dotted lines are unstable states) while the blue
    and orange curves represent asymmetric solutions ($E_1 \neq E_2$). (b) Temporal intensity profile of a single
    PDW connecting the two cw symmetry-broken solutions existing for $X=5$. Blue and orange curves correspond to the
    two polarization modes, while the black curve shows the total power.}
  \label{fig:theo}
\end{figure}

Representative stationary ($\partial/\partial t = 0$) cw (${\partial/\partial\tau = 0}$) solutions of the above
equations are illustrated in Fig.~\ref{fig:theo}(a). Because the equations are symmetric with respect to an
interchange of the two polarization modes, $E_1 \rightleftarrows E_2$, the simplest stationary solutions express this
symmetry ($E_1 = E_2$). Here we retrieve the characteristics S-shaped, bistable response of the Kerr cavity [yellow
curve in Fig.~\ref{fig:theo}(a)]. However, above a certain threshold of driving power, the upper-state solution
undergoes a spontaneous symmetry breaking (SSB): the intensity of the two polarization modes part (blue and orange
curves) \cite{kaplan_directionally_1982, *areshev_polarization_1983, haelterman_polarization_1994,
woodley_universal_2018}. Because of the symmetry of the system, there exists two such solutions, mirror-images of
each other, in which a different mode dominates. These solutions correspond to intracavity fields that are overall
elliptically polarized, with opposite handedness. This polarization SSB has been recently observed
experimentally~\cite{garbin_asymmetric_2019arxiv}. When the two symmetry-broken solutions are simultaneously excited
in different regions (or domains) of the resonator, there exists a shape-preserving temporal structure
interconnecting them and across which the two polarization modes interchange: the PDW [Fig.~\ref{fig:theo}(b)]. Note
how the total intensity (black curve) is nearly constant across the PDW. As shown numerically in
\cite{fatome_polarization_2016}, these dissipative PDWs can circulate indefinitely around the driven resonator
without losing power or changing shape. Their robustness stems from a double balance, similar to that realized for
temporal cavity solitons~\cite{leo_temporal_2010}: the external driving compensates the losses, while dispersive
spreading is balanced by the nonlinearity. The latter occurs through the same mechanism as for the conservative PDWs
described by Haelterman and Sheppard \cite{haelterman_pdw_1994, *sheppard_polarization-domain_1994}.

\begin{figure*}[t]
  \centerline{\includegraphics{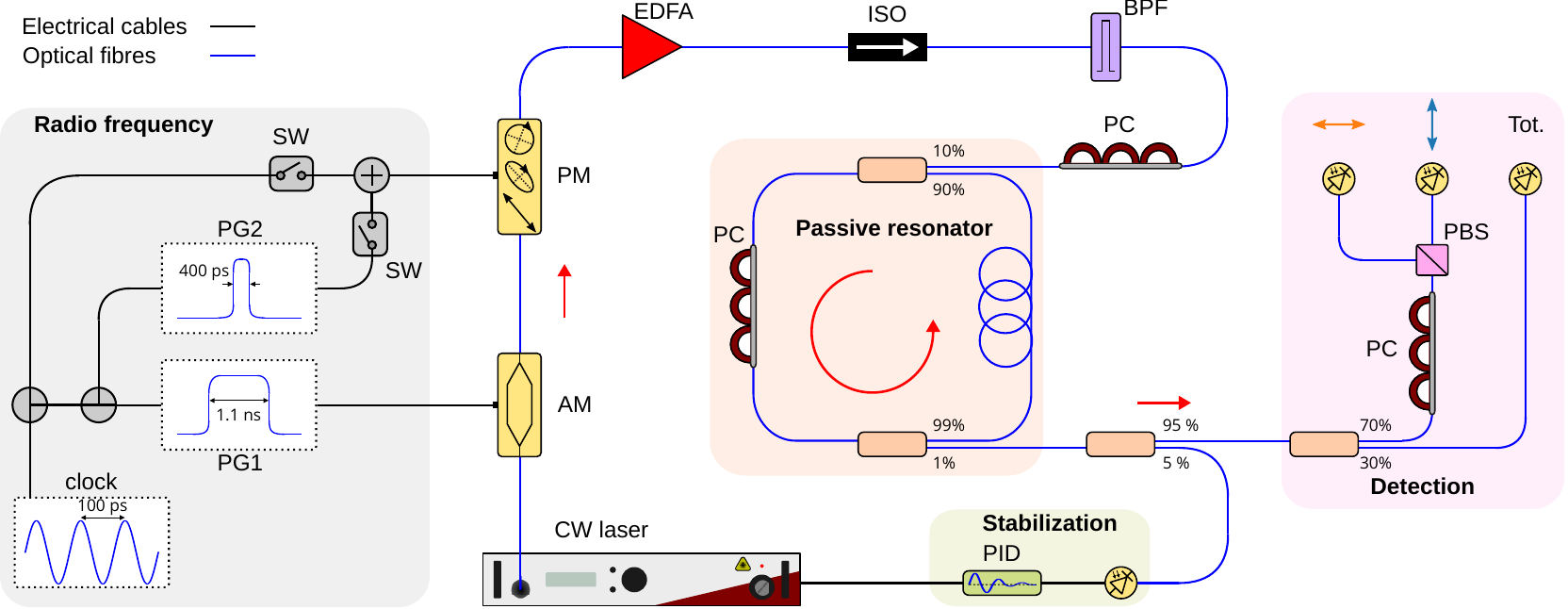}}
  \caption{Experimental set-up. The passive fiber ring resonator is highlighted with an orange background. CW laser,
    driving laser; AM, Mach-Zehnder amplitude modulator; PM, polarization modulator; clock, signal generator; PG1 and
    PG2, pattern generators; SW, electrical switch; EDFA, erbium-doped fibre amplifier; ISO, optical isolator; BPF,
    band-pass filter; PC, polarization controller; PBS, polarizing beamsplitter.}
  \label{fig:setup}
\end{figure*}

The experimental setup that we have used to realize and control dissipative PDWs is depicted in Fig.~\ref{fig:setup}.
It is based on a $\simeq 10$~m-long passive optical fiber ring resonator mostly built out of highly nonlinear, normal
dispersion, ``spun'' fiber, exhibiting very low birefringence due to twisting applied at the drawing stage
\cite{barlow_birefringence_1981}. The ring is closed with two SMF-28 fiber couplers, with splitting ratio 90/10 and
99/1, enabling injection of the driving and monitoring of the intracavity field, respectively. Overall the resonator
exhibits normal dispersion at the 1550-nm~driving wavelength, with averaged second order dispersion coefficient
$\langle\beta_2\rangle \simeq 53\ \mathrm{ps^2/km}$ and nonlinear coefficient $\langle\gamma\rangle \simeq 4.3\
\mathrm{W^{-1}\,km^{-1}}$. The free-spectral-range is found to be $19.8$~MHz, corresponding to a round-trip time
$t_\mathrm{R}$ of~$50.6$~ns. The measured finesse is about~24, amounting to losses of 26\,\% per round-trip, a photon
lifetime of about $4\,t_\mathrm{R}$, and a resonance width of 825~kHz.

The resonator is synchronously driven with flat-top $1.1$~ns pulses~\cite{anderson_observations_2016}. These pulses
are obtained by carving the cw output of a  1~kHz linewidth, erbium-doped distributed-feedback fiber laser with a
10~GHz-bandwidth Mach-Zehnder amplitude modulator (AM). The AM is followed by a fast polarization modulator (PM) used
to apply  perturbations to the driving polarization as explained below. The two modulators are connected to separate
pattern generators (PG) synchronized to the same ${\simeq 10}$~GHz sinusoidal clock, set at a harmonic of the FSR.
Before injection into the resonator, the driving pulses are amplified up to 15~W peak power (corresponding to $X$
values up to~30) using an erbium-doped fiber amplifier (EDFA) combined with a band-pass filter (BPF) for rejection of
amplified spontaneous emission noise. At the output, we monitor separately the power of the two polarization modes,
split by a polarizing beam-splitter (PBS) preceded by a polarization controller (PC), as well as the total output
power. These three signals are measured with a triplet of $12.5$~GHz-bandwidth amplified photodiodes. Additionally, a
small fraction of the total output power is monitored and maintained constant by a PID feedback controller acting on
the driving laser frequency, for stabilization of the detuning with respect to environmental fluctuations.

PDWs require interchange symmetry between the two polarization modes of the resonator. Our optical fiber ring is
however slightly birefringent, due to the couplers, which are not built out of spun fiber, as well as unavoidable
fiber bending. To counterbalance the residual cavity birefringence, a PC is incorporated into the fiber ring. In this
configuration, the polarization modes are associated with states of polarization that evolve around the fiber ring,
and that map onto themselves over one round-trip \cite{coen_experimental_1998}. This  evolution is averaged in the
mean-field model, Eqs.~(\ref{eq:coupledLLEs})~\cite{nielsen_coexistence_2019}. Another PC, inserted before the input
coupler, is used to project the driving field equally onto the two modes, and realize balanced driving conditions.

In practice, the setup is adjusted by observing the resonances of the two polarization modes while scanning the
driving laser frequency. A position of the intracavity PC is found for which, close to a point where the resonances
overlap, their separation can be tuned without affecting their relative amplitudes. With the two resonances slightly
apart, i) the output PC is set to correctly separate the modes in the detection stage, and ii) driving is balanced by
matching the amplitudes of the observed resonances. Birefringence is then cancelled by superimposing the two
resonances. Finally, we increase the driving power until we observe the polarization SSB described in
Fig.~\ref{fig:theo}(a) (and reported in~\cite{garbin_asymmetric_2019arxiv}), and we lock the detuning within the
region where SSB occurs.

\begin{figure*}[t]
  \centerline{\includegraphics[width=2.\columnwidth]{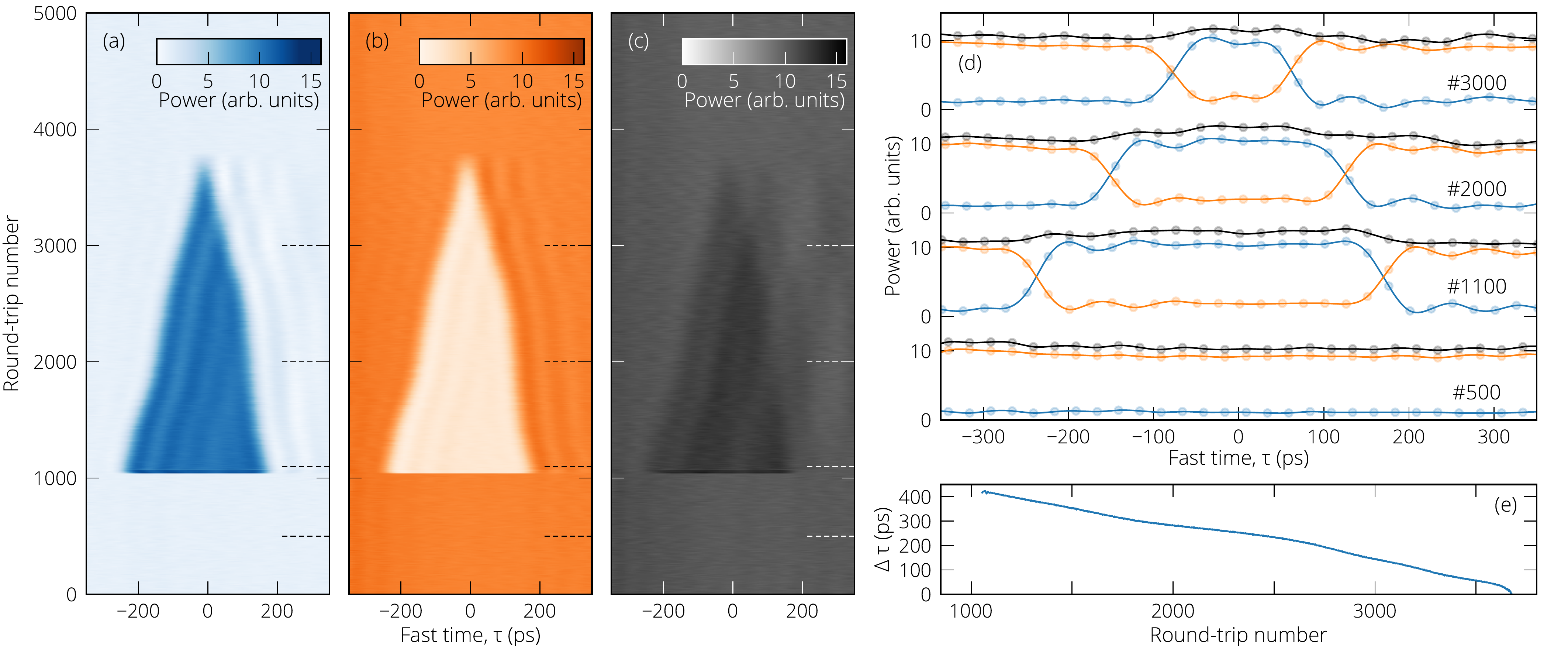}}
  \caption{Experimental evidence of dissipative PDWs for $X \simeq 30.5$ and $\Delta \simeq 13.5$.
    (a)--(c) Round-trip-by-round-trip evolution of the fast-time output power profile for the two polarization
    components, $|E_{1,2}(\tau)|^2$ [(a), blue, and (b), orange] as well as for the total signal [(c), grey] before
    and after a short, localized, 400~ps-long polarization perturbation is applied at about round-trip~\#1000. The
    perturbation generates a ``blue''-mode dominated domain connected to the surrounding regions by two PDWs. The PDWs
    drift towards each other, eventually mutually annihilating. (d) provides corresponding line plots using matching
    colors for selected round-trips as indicated (and marked as dashed side-lines in (a)--(c)]. (e) Evolution of the
    temporal separation~$\Delta\tau$ between the PDWs.}
  \label{fig:pdw}
\end{figure*}

To proceed with observations of PDWs, we record the output power levels across our driving pulses over subsequent
cavity round-trips using a 13-GHz-bandwidth real-time oscilloscope. A typical evolution is shown as color plots
(bottom-to-top) in Figs.~\ref{fig:pdw}(a)--(c), with the three panels corresponding respectively to the powers of the
two separate polarization modes and their total. Using matching colors, line plots are also presented in
Fig.~\ref{fig:pdw}(d) for selected round trips. As can be seen, we start in a symmetry-broken state, where the
``orange'' mode uniformly dominates; see round trip~\#500 in panel~(d). After about 1000 cavity round-trips, a
localized, 400~ps-wide, rf perturbation is applied for about 20~round-trips to the PM. In this way, we carve a domain
of different polarization in the middle of the driving pulse. We then let the intracavity field evolve freely for the
rest of the measurement. Shortly after applying the perturbation, we observe at the output a sudden increase of the
``blue mode'' at a location corresponding to the perturbation, correlated with a depression of the ``orange mode'';
see round trip~\#1100 in panel~(d). We now have a ``domain'' in which the blue mode dominates embedded within the
original orange-dominated state. In that domain, the power levels of the two polarization modes have essentially been
interchanged, reflecting the mirror symmetry of the system. This symmetry can be further appreciated by noting that
the color plots of the two polarization components measured throughout the experiment [Figs.~\ref{fig:pdw}(a)
and~(b)] are essentially negative images of each other. Correspondingly, the total output power
[Fig.~\ref{fig:pdw}(c) and black curves in panel~(d)] reveals little sign of the polarization structure of the
intracavity field. We clearly are in presence of an almost pure polarization dynamics.

We identify the transition regions, along the fast-time~($\tau$) coordinate, where the field switches polarization as
two PDWs of opposite symmetry. The evolution shown in Fig.~\ref{fig:pdw} reveals that these PDWs slowly drift towards
each other (at a rate of about $0.15\ \mathrm{ps}/t_\mathrm{R}$); see also panel~(e) where we plot the temporal
separation between the PDWs vs round-trip number. This results in the shrinkage of the blue-mode dominated domain
created by the polarization perturbation. The PDWs eventually collide and mutually annihilate (around round
trip~\#3800), reverting the system to its initial state. If the interchange symmetry between the polarization modes
was perfect, the PDWs would have no preferred direction of motion and would remain still (if away from other PDWs).
We can therefore attribute the PDWs' motion to the presence of residual asymmetries, favoring one state over the
other~\cite{garbin_asymmetric_2019arxiv}. In particular, from the slight excess power visible in the central domain
[Fig.~\ref{fig:pdw}(c)], we can infer that the blue mode may have been driven slightly stronger than the orange mode.
Nevertheless, the PDWs are very robust: they persist for nearly one thousand photon lifetimes while maintaining their
shape (as far as the 80~ps temporal resolution of our real-time oscilloscope allows us to judge), demonstrating their
dissipative and nonlinearly-localized character.

\begin{figure}[t]
  \centerline{\includegraphics[width=\columnwidth]{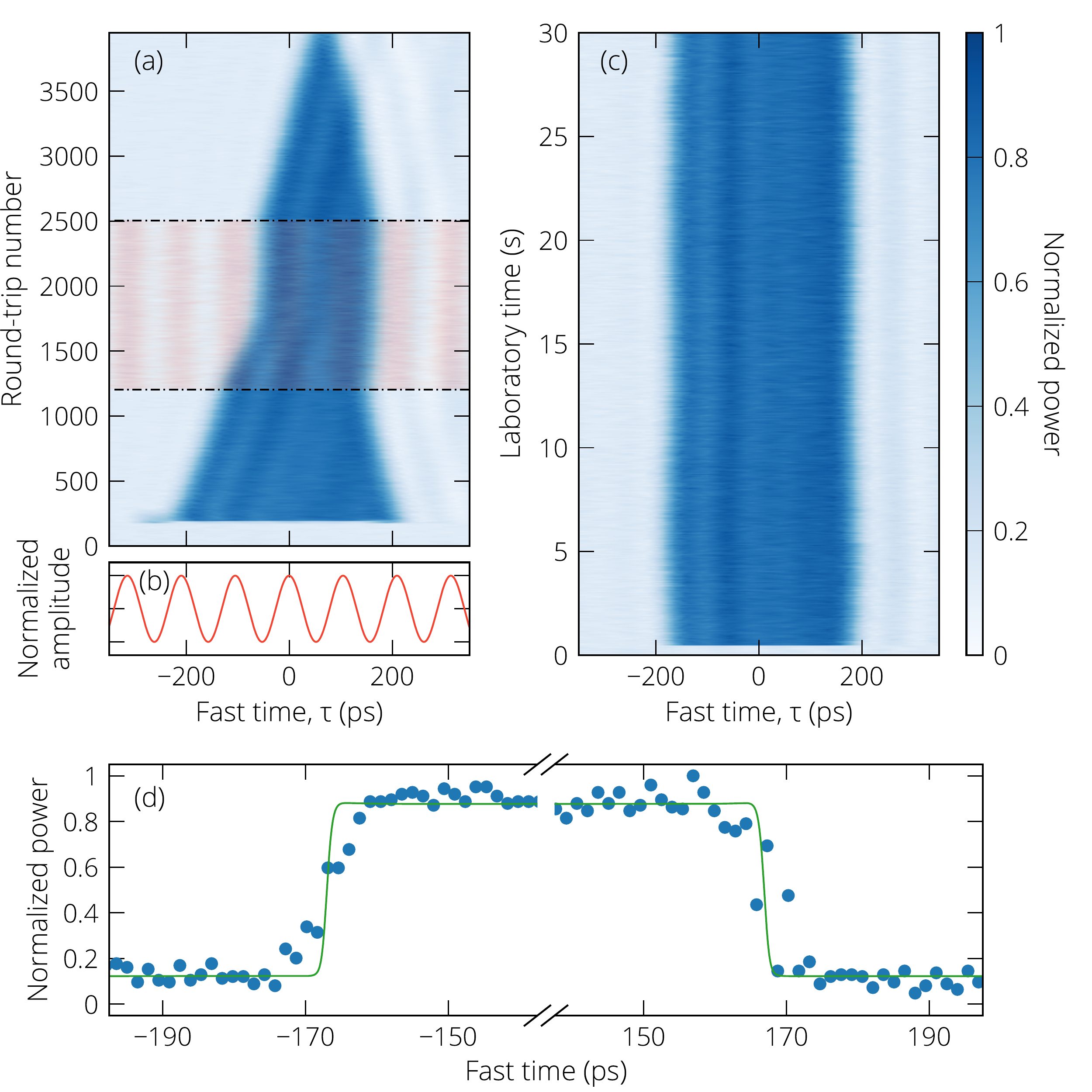}}
  \caption{(a) Demonstration of drifting PDWs being pinned/unpinned to a shallow modulation of the driving
    polarization. We only show the evolution of the output power of one polarization component. The 10~GHz sinusoidal
    modulation [transparent shades of red and panel~(b)] is applied between round-trips~\#1200 and 2500. (c) Long term
    pinning of PDWs, for $X\simeq 24$ and $\Delta \simeq 9.2$. (d) Sampling scope measurement (blue dots) of the
    temporal intensity profile of the two PDWs trapped in~(c). The green curve is the numerical expectation.}
  \label{fig:pin}
\end{figure}

In order to observe stationary PDWs, we have investigated the use of an external modulation of parameters to trap
PDWs, as that technique has been successfully exploited to pin various types of moving fronts in other nonlinear
systems~\cite{rozanov_transverse_1982, *coen_convection_1999, pomeau_front_1986, *marino_front_2014,
haudin_front_2010,  jang_temporal_2015, *garbin_interaction_2017}. In our case, we modulate the polarization of the
driving field, by applying a small fraction of the 10~GHz sinusoidal clock signal to the PM (see
Fig.~\ref{fig:setup}). This modulation can be turned on and off with an additional rf switch (SW).
Figure~\ref{fig:pin}(a) shows the result of an experiment that starts like that discussed in Fig.~\ref{fig:pdw} (only
showing one polarization component), with two PDWs initially drifting towards each other at a constant speed. When
the modulation is turned on at   round-trip~\#1200 [see Fig.~\ref{fig:pin}(b) as well as  the red-shaded area in
Fig.~\ref{fig:pin}(a)], we immediately observe a change of behavior. The PDWs visibly change their drift velocities,
and after some transient, eventually reach a fixed position with respect to the modulation. In that position, the PDW
drift imparted by the local driving imbalance associated with the modulation counteracts the original motion due to
the residual asymmetries. The PDWs hold their position until the modulation is turned off at round-trip~\#2500, which
releases them, back on their original collision course. In Fig.~\ref{fig:pin}(c), using the same technique, we
demonstrate long term pinning of two PDWs over 30~seconds (corresponding to a propagation distance of $6\times
10^6$~km inside the resonator), which has enabled us to measure their temporal intensity profile with a 65-GHz
sampling oscilloscope [blue dots in Fig.~\ref{fig:pin}(d)]. This measurement demonstrates that our PDWs have a rise
time (10--90\,\%) of less than~9~ps, limited by the bandwidth of our oscilloscope, and compatible with the $1.6$~ps
numerical expectation [green curves in Fig.~\ref{fig:pin}(d)].

We remark that the PDWs reported in Figs.~\ref{fig:pdw} and~\ref{fig:pin} are observed for driving powers
comparatively larger than that considered in the theoretical plot of Fig.~\ref{fig:theo}. Large driving powers are
made necessary by the presence of a small amount of linear coupling between the polarization modes of our fiber
resonator. Numerical calculations indicate that linear coupling, which splits the cavity resonance
\cite{pierce_coupling_1954}, thwarts polarization SSB at low power. SSB and PDWs are restored at high power, when the
Kerr-induced tilt of the cavity resonance dominates over the splitting~\cite{del_bino_symmetry_2017}.

In conclusion, we have reported here the first experimental demonstration of dissipative PDWs. The PDWs are
recirculated in a passive, driven Kerr optical fiber ring resonator. Their existence relies on a symmetry breaking
bifurcation and on an interchange symmetry between the two polarization modes of the resonator. Our dissipative PDWs
are found to be robust with respect to residual imperfections and asymmetries, and can be pinned to a shallow
external modulation. Given their duration, our resonator could hold up to 20,000~PDWs in a cw-driven configuration,
which could be achieved by mitigating linear mode coupling as in~\cite{garbin_asymmetric_2019arxiv}, thus reducing
power requirements. Our results suggest that  our system could be used as an all-optical buffer for PDW-based
topological bit transmissions~\cite{gilles_polarization_2017}. Optical PDWs could also prove useful for the real time
stochastic analog simulation of other DW-related phenomena.

\begin{acknowledgments}
  We thank Y.~Wang for technical help, and F.~Leo for fruitful discussions. We acknowledge financial support from The
  Royal Society of New Zealand, in the form of Marsden Funding (18-UOA-310), as well as James Cook (JCF-UOA1701, for
  S.C.) and Rutherford Discovery (RDF-15-UOA-015, for M.E.) Fellowships. J.F. thanks the Conseil r\'egional de
  Bourgogne Franche-Comt\'e, mobility (2019-7-10614).
\end{acknowledgments}


%

\end{document}